\documentclass[aps,prl,superscriptaddress,twocolumn,floatfix,showpacs,amsmath,amssymb]{revtex4}

\usepackage{graphicx}
\usepackage{amsmath}
\usepackage{amssymb}
\usepackage{color}
\usepackage{dcolumn}
\usepackage{epsfig}
\usepackage{bm}
\usepackage[urlcolor=blue]{hyperref}
\hypersetup{backref, colorlinks=true, linkcolor=blue, citecolor=blue}

\begin{document}
\title{Superconductivity in Dense Rashba Semiconductor BiTeCl}

\author{Jian-Jun Ying}
\affiliation{Center for High Pressure Science and Technology Advanced Research, Shanghai 201203, China}
\affiliation{Geophysical Laboratory, Carnegie Institution of Washington, Washington, DC 20015, U.S.A.}

\author{Viktor V. Struzhkin}
\affiliation{Geophysical Laboratory, Carnegie Institution of Washington, Washington, DC 20015, U.S.A.}

\author{Alexander F. Goncharov}
\affiliation{Geophysical Laboratory, Carnegie Institution of Washington, Washington, DC 20015, U.S.A.}

\author{Ho-Kwang Mao}
\affiliation{Geophysical Laboratory, Carnegie Institution of Washington, Washington, DC 20015, U.S.A.}
\affiliation{Center for High Pressure Science and Technology Advanced Research, Shanghai 201203, China}

\author{Fei Chen}
\affiliation{Hefei National Laboratory for Physical Sciences at Microscale and Key Laboratory of Strongly-Coupled Quantum Matter Physics, Chinese Academy of Sciences, School of Physical Sciences, University of Science and Technology of China, Hefei 230026, China}
\affiliation{Center for High Pressure Science and Technology Advanced Research, Shanghai 201203, China}

\author{Xian-Hui Chen}
\affiliation{Hefei National Laboratory for Physical Sciences at Microscale and Key Laboratory of Strongly-Coupled Quantum Matter Physics, Chinese Academy of Sciences, School of Physical Sciences, University of Science and Technology of China, Hefei 230026, China}
\affiliation{Collaborative Innovation Center of Advanced Microstructures,  Nanjing 210093, China}

\author{Alexander G. Gavriliuk}
\affiliation{Institute of Crystallography, Russian Academy of Sciences, Leninsky pr. 59, Moscow 119333, Russia}
\affiliation{Institute for Nuclear Research, Russian Academy of Sciences, Troitsk, Moscow 142190, Russia}

\author{Xiao-Jia Chen}
\email{xjchen@hpstar.ac.cn}
\affiliation{Center for High Pressure Science and Technology Advanced Research, Shanghai 201203, China}
\affiliation{Geophysical Laboratory, Carnegie Institution of Washington, Washington, DC 20015, U.S.A.}

\date{\today}

\begin{abstract}
Layered non-centrosymmetric bismuth tellurohalides are being examined as candidates for topological insulators. Pressure is believed to be essential for inducing and tuning topological order in these systems. Through electrical transport and Raman scattering measurements, we find superconductivity in two high-pressure phases of BiTeCl with the different normal state features, carrier characteristics, and upper critical field behaviors. Superconductivity emerges when the resistivity maximum or charge density wave is suppressed by the applied pressure and then persists till the highest pressure of 51 GPa measured. The huge enhancement of the resistivity with three magnitude of orders indicates the possible achievement of the topological order in the dense insulating phase. These findings not only enrich the superconducting family from topological insulators but also pave the road on the search of topological superconductivity in bismuth tellurohalides.
\end{abstract}


\vskip 300 pt

\maketitle Topological insulators represent the newly discovered phase of matter with insulating bulk state but topologically protected metallic surface state due to the time-reversal symmetry and strong spin-orbital interaction \cite{Hasan,XLQi1}. Searching for topological superconductivity is one of the hottest topics due to the exploration of fundamental physics and the potential applications in topological quantum computation \cite{Nayak, Alicea}. Superconductivity has been found in some topological insulators such as compressed A$_2$B$_3$ compounds including Bi$_2$Te$_3$ \cite{MAIlina,Einaga,JLZhang,CZhang,Matsubayashi}, Bi$_2$Se$_3$ \cite{kong1,Kirshenbaum}, Sb$_2$Te$_3$ \cite{JZhu}, and Sb$_{2}$Se$_{3}$ \cite{kong} and substituted Cu$_x$Bi$_2$Se$_3$ \cite{Hor} together with related materials such as YPtBi \cite{Butch}. However, the identification of their topological superconductivity is still a hard task and under debate \cite{JLZhang,Matsubayashi}. In most cases, pressure is needed to drive topological insulators to superconductors. Superconductivity is usually accompanied by the electronic topological transition and/or structural transition \cite{Einaga,Kirshenbaum}. It remains unclear whether such a transition is essential for inducing superconductivity in topological insulators.

The class of non-centrosymmetric bismuth tellurohalides (BiTeX with X=Cl, Br, I) exhibit large Rashba-type splittings in the bulk bands \cite{Ishizaka, Eremeev,Crepaldi, Landolt, Landolt2}, and they are potential candidates for building the spintronic devices. Pressure-induced topological quantum phase transition was predicted for Rashba semiconductor BiTeI \cite{Bahramy}. However, controversial conclusions were drawn from the following experiments on this material \cite{Xi,Tran}. Recently, BiTeCl was discovered to be the first example of inversion asymmetric topological insulator (IATI) from angle-resolved photoemission spectroscopy (ARPES) experiment \cite{YLChen}. This was soon  supported by the transport measurement \cite{FXXiang}. Unlike the previously discovered three dimensional topological insulators with inversion symmetry, the inversion symmetry is naturally broken by the crystal structure in IATI. It is highly possible  to realize the topological magneto-electric effects and the topological superconductivity \cite{YLChen,LFu,XLQi}. However, quantum oscillation measurements excluded the existence of Dirac surface state in BiTeCl single crystals \cite{FChen, Martin}. Such contradiction may come from the strong surface polarity which would generate large effective pressure along the $c$ axis. This pressure could drive the several surface layers into a topological insulator as the case in BiTeI \cite{Bahramy,YLChen}. Therefore, pressure must be used in experiments to examine whether topological phase transition could happen in this kind of materials and whether superconductivity would be induced after that. Meanwhile, the discoveries of pressure-induced superconductivity from topological insulators were limited in A$_2$B$_3$-type compounds. Finding new superconducting family from topological insulators would add a new opportunity for exploring topological superconductivity.

In this Letter, we address the above mentioned issues by investigating the high-pressure behaviors of BiTeCl through the combination of the electrical transport properties and Raman scattering measurements. A huge enhancement of the resistivity with three magnitude of orders is observed with the application of pressure followed by a phase transition at relatively low pressure of 5 GPa, pointing to the possible realization of topological order in the insulating phase suggested from recent experiment. Two following superconducting phases are discovered with the highest critical temperature $T_{c}$ of 7 K at 15 GPa. These results indicate that high-pressure study offers a new opportunity for uncovering the novel physical properties in bismuth tellurohalides.

High quality single crystals of BiTeCl were grown by a self-flux method \cite{Landolt2}. Pressure was applied at room temperature using the miniature diamond anvil cell \cite{Gavriliuk}. Diamond anvil with 300 $\mu$m culet and $c$-BN gasket with sample chambers of diameter 130 $\mu$m were used. BiTeCl single crystal was cut with the dimensions of 65$\times$65$\times$15 $\mu$m$^{3}$. Four Pt wires were adhered to the sample using the silver epoxy. Daphne oil 7373 was used as a pressure transmitting medium. Pressure was calibrated by using the ruby fluoresce shift at room temperature. Resistivity and Hall coefficient were measured using the Quantum Design PPMS-9. Diamond anvil with 300 $\mu$m culet was used for high-pressure Raman measurements with incident laser wavelength of 488 nm. Neon was loaded as the pressure transmitting medium.

\begin{figure}[t]
\centering
\includegraphics[width=\columnwidth]{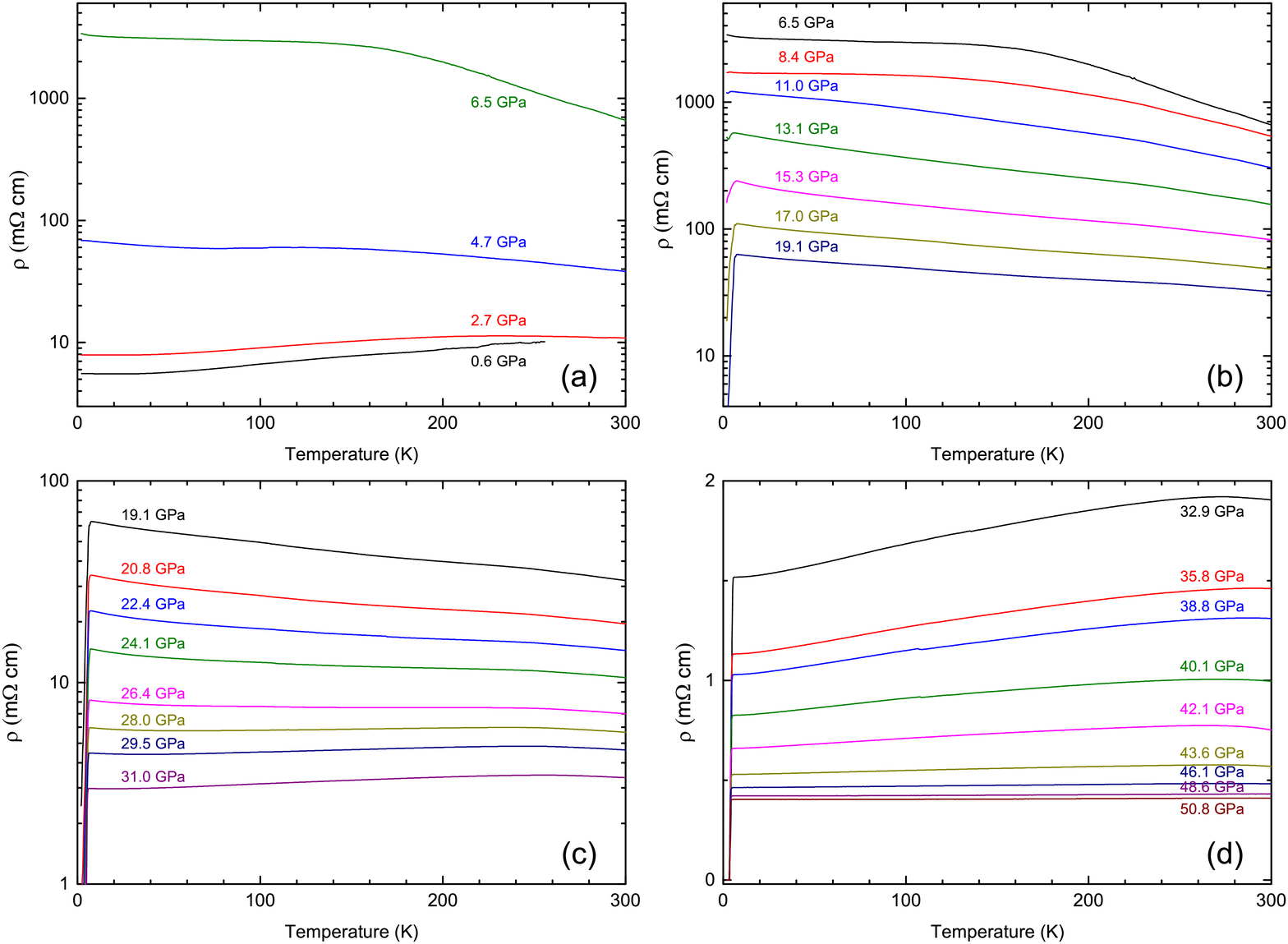}
\caption{(Color online) Temperature dependence of the resistivity of BiTeCl at various pressures. (a) The sudden change of the resistivity behavior indicates a phase transition from Rashba semiconductor to insulator around 5 GPa. (b) The resistivity was gradually suppressed with increasing pressure and superconductivity emerges above 13 GPa. The normal state behaves like an insulator (c) and a metal (d) upon heavy compression. }
\end{figure}

\begin{figure}[t]
\centering
\includegraphics[width=\columnwidth]{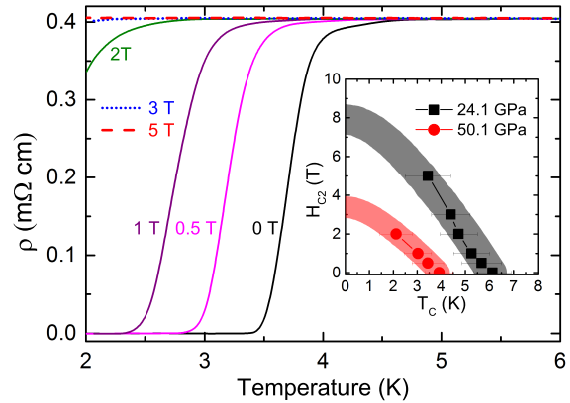}
\caption{(Color online) Temperature dependence of the resistivity of BiTeCl at various magnetic fields and pressure of 50.8 GPa. Inset: Upper critical field $H_{\rm c2}$ for the pressure of 24.1 GPa and 50.8 GPa, respectively. $T_{c}$ was determined from the 90\% resistivity transition. The color area represents the calculated $H_{\rm c2}$ from WHH theory.}
\end{figure}

\begin{figure}[t]
\centering
\includegraphics[width=0.5\textwidth]{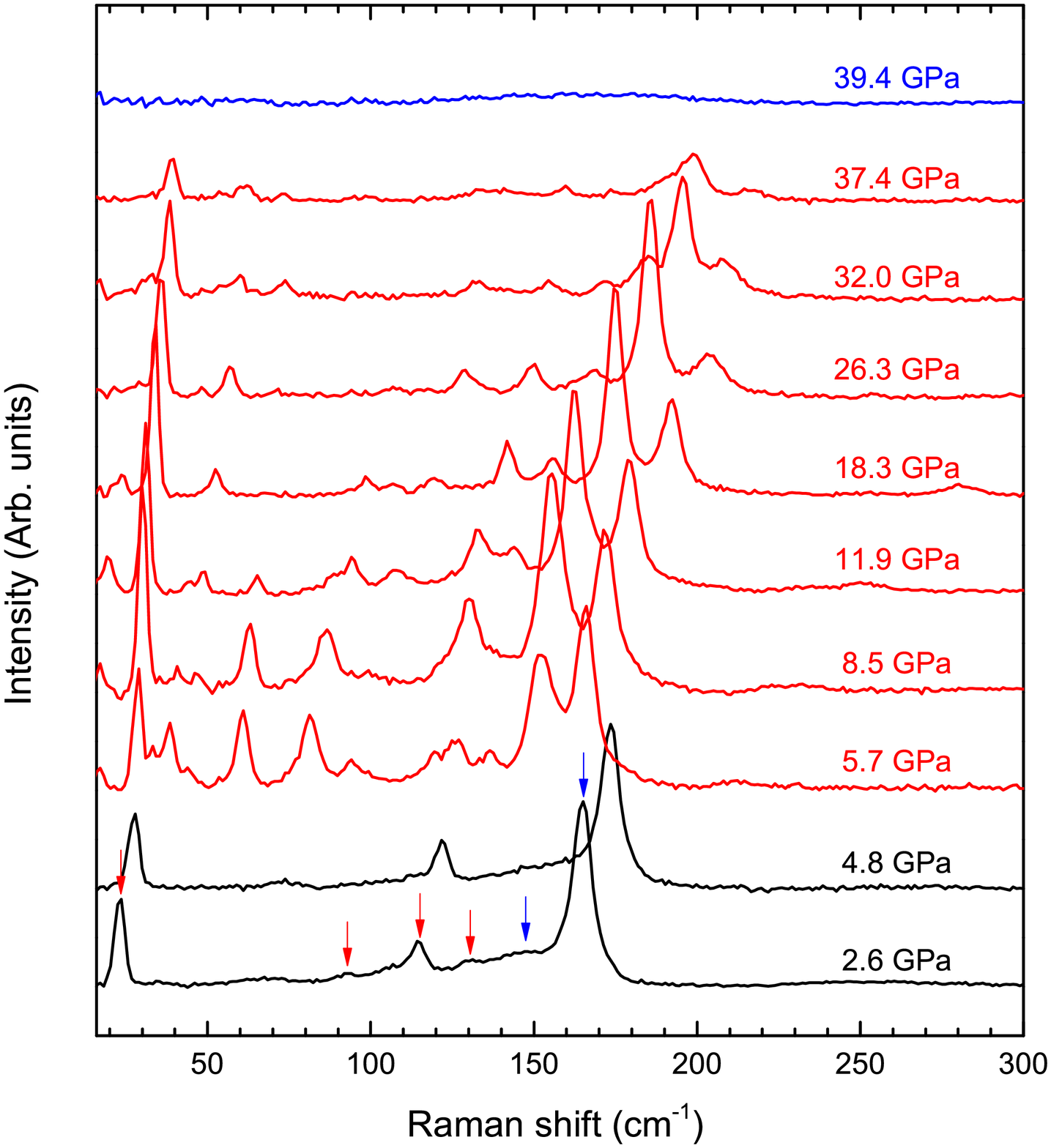}
\caption{(Color online) Raman spectra of BiTeCl measured under pressure. The vibration modes in phase I are marked by arrows. The different spectra above 5 GPa and the absence of Raman peaks above 39 GPa indicate two different phases.}
\end{figure}

\begin{figure}[t]
\centering
\includegraphics[width=0.5\textwidth]{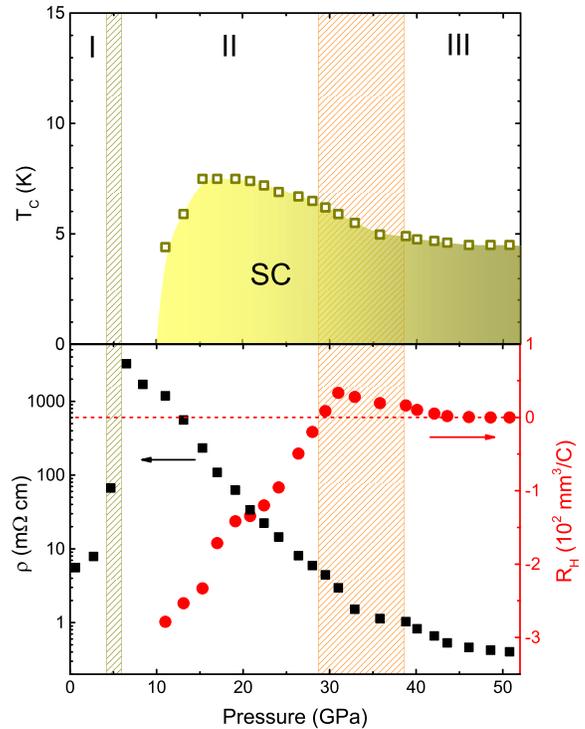}
\caption{(Color online) Phase diagram of BiTeCl under pressure (upper panel) and pressure dependence of the resistivity and Hall coefficient measured at 10 K (lower panel). The dashed regions represent the borders of two phase transitions. Two high-pressure phases have large overlapping region. }
\end{figure}

Figure 1 shows the temperature dependence of the resistivity for BiTeCl at various pressures up to 50.8 GPa. At low pressures below 4 GPa, the resistivity shows a metallic behavior similar to that at ambient pressure as reported before \cite{TIdeue,Akrap}, though this material was thought to be a Rashba semiconductor. When pressure is increased to 5 GPa, the resistivity suddenly increases almost 3 orders, and shows an insulating behavior with a resistivity maximum around 150 K [Fig. 1(a)]. It clearly indicates a phase transition from a Rashba semiconductor to an insulator around 5 GPa in BiTeCl. Further increasing pressure, the resistivity is gradually reduced and the resistivity maximum shifts downwards to lower temperatures. Interestingly, superconductivity emerges when the resistivity maximum is completely suppressed around 13 GPa, though the normal state still exhibits an insulating behavior. This behavior is different from the other superconducting phases obtained from topological insulators under pressure \cite{Einaga, Kirshenbaum}. Further increasing pressure, the normal state behaves gradually in a metallic way above 28 GPa. Superconductivity persists up to  the highest pressure of 50.8 GPa studied. Superconductivity in compressed BiTeCl has never been reported before. Such a discovery with novel normal-state properties in BiTeCl enriches the physics besides large Rashba-type splittings and topological properties.

The obtained superconductivity in BiTeCl was further supported by the evolution of the resistivity-temperature curve with the applied magnetic fields (Fig. 2). The curve gradually shifts towards the lower temperatures with increasing magnetic field. The magnetic field was applied along the crystallographic $c$ axis. It seems likely that 5 T is sufficient to suppress superconductivity at 50.8 GPa. However, much higher field is needed to suppress superconductivity at 24.1 GPa. These two pressures yield different normal-state behaviors $-$ metallic for the former but insulating for the latter (Fig. 1). The upper critical field $H_{\rm c2}$ can be determined from these measurements.

Within the weak-coupling BCS theory, the upper critical field at T = 0 K can be determined by the Werthamer-Helfand-Hohenberg (WHH) equation \cite{WHH}: $H_{c2}(0)=0.693[-(dH_{c2}/dT)]_{T_{c}}T_{c}$. The grey and red areas shown in the inset of Fig. 2 are temperature dependence of $H_{\rm c2}$ calculated based on the WHH theory for the superconductivity at 24.1 and 50.8 GPa, respectively. The $H_{\rm c2}(0)$ at 50.8 GPa is about 3.5 T. This value is comparable with that of superconducting Bi$_2$Se$_3$ phase \cite{Kirshenbaum}. The calculated $H_{\rm c2}(0)$ of almost 8 T at 24.1 GPa is twice larger than that at 50.8 GPa. The large difference of $H_{\rm c2}(0)$ at 24.1 and 50.8 GPa indicates the different origins of superconductivity of two high-pressure phases. Differing from A$_2$B$_3$-type topological materials, the normal state of superconducting BiTeCl shows both insulating and metallic behaviors with different values of $H_{\rm c2}(0)$.

Raman spectroscopy is a very powerful tool to probe the changes in lattice and thus can provide valuable information on structure. Figure 3 shows the Raman spectra of BiTeCl at various pressures up to 39.4 GPa. At ambient pressure, BiTeCl crystalizes in a trigonal layer structure with space group of $P6_3mc$ (phase I) \cite{Shevelkov}. This phase has seven Raman active modes (2$A_{1}$+2$E_{1}$+3$E_{2}$). Both $E$-type modes belong to the in-plane vibration of the Bi, Te, and Cl layers, while the lower $E$ mode has a large contribution of the Cl atom vibration. The two out-of-plane $A_{1}$ modes include the vibration with higher frequencies. Our Raman measurements produced all these modes in phase I below 5 GPa. The obtained modes marked by the arrows are also similar to those reported previously at ambient pressure \cite{IYu, Akrap}.

Above 5 GPa, the spectra suddenly change, indicating the appearance of a new high-pressure phase. The similar phase transition to an orthorhombic $Pnma$ structure has been reported in the sister system BiTeI \cite{YChen}. The $Pnma$ structure has more Raman vibrational modes (6$A_{g}$+3$B_{1g}$+6$B_{2g}$+3$B_{3g}$). This phase was predicted to be a semiconductor in BiTeI \cite{YChen}. The temperature dependence of the resistivity of BiTeCl shown in Fig. 1(b) exhibits a more complicated feature. There is a maximum around 150 K for the pressure of 6.5 GPa. Applying pressure shifts it down to lower temperatures. However, the maximum feature is suppressed when superconductivity appears by applied pressure. The emergence of superconductivity and the destruction of the resistivity maximum at the same pressure indicate a close connection between them. The resistivity maximum is the common feature for charge density wave. This superconducting phase is possible to come after the suppression of the charge density wave. The evolution of charge density wave with pressure has been observed in 1$T$-TiSe$_{2}$ \cite{joe}.

Above 39 GPa, no Raman modes were detected, suggesting a phase transition to a high symmetric structure with a possible cubic unit cell. This is consistent with the obtained metallic normal state of this high-pressure phase (phase III) from the resistivity measurements [Fig. 1(d)]. Previously, an orthorhombic $P$4/$nmm$ structure has been suggested for phase III of BiTeI from combined XRD measurements and theoretical calculations \cite{YChen}. This $P$4/$nmm$ structure corresponds to more rich Raman active modes with nine $E_{g}$ modes besides $A_{1g}$, $B_{1g}$, and $B_{2g}$. Our Raman data does not support the existence of the $P$4/$nmm$ structure for BiTeCl. The metallic phase of BiTeCl is possibly a substitutional alloy similar to Bi$_2$Te$_3$ system \cite{Zhu}. The Raman spectra provide evidence for the existence of two phase transitions and the features of different phases in BiTeCl.

Combining the resistivity and Raman measurements, we can map out the phase diagram of BiTeCl (Fig. 4). Below 5 GPa, BiTeCl keeps its initial phase I as a Rashba semiconductor but behaves in reality like a metal or semimetal [Fig. 1(a)]. Above that, it evolves to a superconductor with non-metallic normal state in phase II up to the phase boundary starting at 28 GPa. Above 39 GPa, superconductivity in phase III remains unchanged with pressure but the material possesses a metallic normal state. $T_{c}$ exhibits a dome-like shape with pressure in phase II, while it becomes almost constant upon heavy compression in phase III. Accompanying the $T_{c}$ change, the sign of the measured Hall coefficient at 10 K gradually changes from negative to positive above 28 GPa (lower panel of Fig. 4). This provides the dominant electron and hole carrier character for the superconducting phase II and III. The large hole carrier density in phase III is consistent with the character of the substitutional alloy indicated from Raman measurement. The relatively weak pressure dependence of the carrier density provides a natural explanation for the almost constant $T_{c}$ in this phase.

When examining the electrical transport properties at different high-pressure phases of BiTeCl, one can see that the so-called Rashba semiconducting phase at ambient pressure behaves like a metal \cite{TIdeue,Akrap} or a semimetal due to the apperrance of the resistivity maximum [Fig. 1(a)]. Superconductivity emerges when the resistivity maximum or charge density wave is suppressed by applied pressure. The superconducting phase II  possesses an insulating or a semiconducting normal state [Figs. 1(b) and 1(c)]. This dramatic change of the electrical transport with pressure is also reflected by the resistivity value. The lower panel of Fig. 4 also summarizes the resistivity measured at 10 K. An anomaly is clearly observed across the phase boundary of phase I and II. The semiconductor-semimetal transition serves a possibility for an excitonic insulator. The crossover of the carrier character again indicates a possibility of excitonic superconductivity due to Coulomb attraction of electrons and holes and the formation of Bose-condensate of the electron-hole pairs. For all the features observed, particularly for the evolution of the resistivity maximum with pressure and the emergence of superconductivity, BiTeCl is considered as a new excitonic insulator and excitonic superconductor. Meanwhile, the Rashba semiconducting phase at ambient pressure can be tuned into an insulating phase above 5 GPa. The huge enhancement of the resistivity with three magnitude of orders in phase I indicates the possible realization of the topological insulator upon compression suggested from early experiment. Although the estimated pressure of the surface polarity along the $c$ axis for BiTeCl is only about 1 GPa in the ARPES measurement \cite{YLChen}, our measurements were performed at the quasi-hydrostatic pressure condition which is quite different with the nonhydrostatic environment in the ARPES experiment. The pressure-induced topological phase transition at 4-5 GPa has also been observed in the sister BiTeI \cite{YChen,yus}. These results together contribute BiTeCl the possible topological feature in its insulating state and even superconducting state.

In conclusion, we have reported an experimental finding of superconductivity in dense Rashba semiconductor BiTeCl. Superconductivity emerges after the suppression of the resistivity maximum or charge density wave around 13 GPa and persists above 51 GPa. The interesting electrical transport properties and the crossover of the carrier characteristics upon compression make BiTeCl to be the candidate for a new excitonic insulator and excitonic superconductor. The significant enhancement of the resistivity in the insulating phase indicates the possible realization of the topological order. These results demonstrate that BiTeCl possesses novel physical properties under pressure besides the extraordinary large Rashba-type splittings and topological properties. Our work thus paves the road on searching topological superconductivity in BiTeX (X=Cl, Br, I) compounds.

\begin{acknowledgments}
This work was supported by EFree, an Energy Frontier Research Center funded by the U.S. Department of Energy (DOE), Office of Science, Office of Basic Energy Sciences (BES) under Award Number DE-SG0001057. The resistance measurements were supported by the DOE under Grant No. DE-FG02-02ER45955. Raman measurements were supported by the U.S. National Science Foundation Earth Sciences Instrumentation and Facilities (EAR/IF) and DARPA. The sample design and growth were supported by the National Natural Science Foundation of China (Grant No. 11190021), the ``Strategic Priority Research Program (B)" of the Chinese Academy of Sciences (Grant No. XDB04040100). A.G.G. acknowledges the support from Russian Foundation for Basic Research (Grant No. 14-02-00483-a), Russian Scientific Foundation (Grant No. 14-12-00848), and ``Elementary particle physics, fundamental nuclear physics and nuclear technologies''.
\end{acknowledgments}

\end{document}